\documentclass[3p,times,authoryear]{elsarticle}

\usepackage{amssymb}
\usepackage[colorlinks]{hyperref}
\usepackage{threeparttable}

\journal{}

\begin{document}

\begin{frontmatter}

%% Title, authors and addresses

%% use the tnoteref command within \title for footnotes;
%% use the tnotetext command for theassociated footnote;
%% use the fnref command within \author or \address for footnotes;
%% use the fntext command for theassociated footnote;
%% use the corref command within \author for corresponding author footnotes;
%% use the cortext command for theassociated footnote;
%% use the ead command for the email address,
%% and the form \ead[url] for the home page:
%% \title{Title\tnoteref{label1}}
%% \tnotetext[label1]{}
%% \author{Name\corref{cor1}\fnref{label2}}
%% \ead{email address}
%% \ead[url]{home page}
%% \fntext[label2]{}
%% \cortext[cor1]{}
%% \address{Address\fnref{label3}}
%% \fntext[label3]{}

\title{Chirality and magnetic configuration associated with two-ribbon solar flares: \\AR 10930 versus AR 11158}

%% use optional labels to link authors explicitly to addresses:
%% \author[label1,label2]{}
%% \address[label1]{}
%% \address[label2]{}

\author[naoc,sdu,ucas]{Han He\corref{cor1}}
\cortext[cor1]{Corresponding author.}
\ead{hehan@nao.cas.cn}
\author[naoc,ucas]{Huaning Wang}
\author[naoc,ucas]{Yihua Yan}
\author[sdu]{Bo Li}
\author[nju,klnju]{P. F. Chen}

\address[naoc]{CAS Key Laboratory of Solar Activity, National Astronomical Observatories, Chinese Academy of Sciences, 20A Datun Road, Chaoyang District, Beijing 100101, China}
\address[sdu]{Shandong Provincial Key Laboratory of Optical Astronomy and Solar-Terrestrial Environment, Institute of Space Sciences, Shandong University, 180 Wenhua Xilu, Weihai 264209, China}
\address[ucas]{University of Chinese Academy of Sciences, Beijing, China}
\address[nju]{School of Astronomy and Space Science, Nanjing University, 163 Xianlin Street, Qixia Distirct, Nanjing 210023, China}
\address[klnju]{Key Laboratory for Modern Astronomy and Astrophysics (Nanjing University), Ministry of Education, Nanjing 210023, China}

\begin{abstract}
  The structural property of the magnetic field in flare-bearing solar active regions (ARs) is one of the key aspects for understanding and forecasting solar flares. In this paper, we make a comparative analysis on the chirality and magnetic configurations associated with two X-class two-ribbon flares happening in AR 10930 and AR 11158. The photospheric magnetic fields of the two ARs were observed by space-based instruments, and the corresponding coronal magnetic fields were calculated based on the nonlinear force-free field model. The analysis shows that the electric current in the two ARs was distributed mostly around the main polarity inversion lines (PILs) where the flares happened, and the magnetic chirality (indicated by the signs of force-free factor $\alpha$) along the main PILs is opposite for the two ARs, i.e., left-handed ($\alpha<0$) for AR 10930 and right-handed ($\alpha>0$) for AR 11158. It is found that, for both the flare events, a prominent magnetic connectivity (featured by co-localized strong $\alpha$ and strong current density distributions) was formed along the main PIL before flare and was totally broken after flare eruption. The two branches of the broken magnetic connectivity, combined with the prominent magnetic connectivity before flare, compose the opposite magnetic configurations in the two ARs owing to their opposite chirality, i.e., Z-shaped configuration in AR 10930 with left-handed chirality and inverse Z-shaped configuration in AR 11158 with right-handed chirality. It is speculated that two-ribbon flares can be generally classified to these two magnetic configurations by chirality in the flare source regions of ARs.
\end{abstract}

%%Graphical abstract
%\begin{graphicalabstract}
%\includegraphics{grabs}
%\end{graphicalabstract}

%%Research highlights
%\begin{highlights}
%\item Research highlight 1
%\item Research highlight 2
%\end{highlights}


\end{frontmatter}

%\linenumbers

%% main text
\section{Introduction}\label{sec:intro}

Solar flare is a significant eruptive phenomenon observed in the solar atmosphere, which manifests as sudden enhancement across a broad band of electromagnetic wave spectrum (e.g., optical emission in photosphere and chromosphere, and soft X-ray emission in corona). It is believed to be the result of magnetic energy release in corona through the process of magnetic reconnection \citep{2002A&ARv..10..313P, 2011SSRv..159...19F, 2011SSRv..158....5H, 2011LRSP....8....6S, 2017LRSP...14....2B}. A typical major solar flare event is usually accompanied with coronal mass ejection (CME) \citep{2000JGR...10523153F, 2001ApJ...559..452Z, 2005ARA&A..43..103Z, 2011LRSP....8....1C, 2012LRSP....9....3W, 2016GSL.....3....8G}. Beneath the erupting plasmas is the flaring arcade. At the footpoints of the flaring arcade, two bright flare ribbons can be seen in chromospheric images observed via chromospheric spectral lines such as H$\alpha$, Ca \textsc{ii} H\&K, etc. This sort of typical solar flare events presenting two flare ribbons are also referred to as two-ribbon flares (see example images in Figure \ref{fig:fig1}).

Big flares and associated CMEs can cause drastic disturbances to the solar-terrestrial and heliospheric space weather conditions \citep[e.g.,][]{2006LRSP....3....2S, 2007LRSP....4....1P, 2011JGRA..11611301L, 2013Sci...341.1489G, 2016NatSR...632362H, 2019NatAs...3.1024G, 10.3389/fphy.2020.00045} and impact the safety of spacecrafts as well as the human society on the Earth \citep[e.g.,][]{2007LRSP....4....1P, 2017SSRv..212.1253L}. Similar phenomenon of flare activity was also discovered on the solar-type stars other than the Sun \citep[e.g.,][]{2000ApJ...529.1026S, 2012Natur.485..478M, 2013ApJ...771..127N, 2013ApJS..209....5S, 2015MNRAS.447.2714B, 2016ApJ...829...23D}. Understanding the physics of solar flares will help to improve the predictions of severe space weather events \citep[e.g.,][]{2006LRSP....3....2S, 2017SSRv..212.1137K}, and will also be helpful for investigating the process of stellar activities \citep[e.g.,][]{2010ARA&A..48..241B, 2011LRSP....8....6S, 2013A&A...549A..66A, 2017EPJWC.16002009B, 2017SCPMA..60a9601C, 2018ApJS..236...19E, 2018SSRv..214...45M}.

Solar flares are usually observed in solar active regions (ARs) which are sunspot complexes, and the existence of sunspots indicates the strong and concentrated magnetic fields within solar ARs. The structural property of the magnetic field in flare-bearing ARs is one key aspect for understanding the physical processes of solar flares and other related eruptive phenomena in the solar atmosphere \citep[e.g.,][]{2012ChSBu..57.1393C, 2012NatCo...3E.747Z, 2014JGRA..119.3286H, 2015ApJ...802L..21K, 2015NatCo...6E7598S, 2016NatCo...711522J, 2016NatSR...634021L, 2017ScChE..60.1383C, 2017ScChE..60.1408G, 2017NatCo...8.2202H, 2017NatCo...8.1330W, 2017ApJ...834..150Y, 2018ApJ...867...83I, 2018ApJ...869...13J, 2018SCPMA..61f9611L, 2019ApJ...874..181G}. The nonpotential property of the magnetic fields with electric current permeating inside is believed to be the key aspect related to the activity level of ARs \citep[e.g.,][]{1996ApJ...456..861W, 2003ApJ...595.1296L, 2016ApJ...820..103S}.

In the standard model of solar flares (also known as CSHKP model; see e.g., \citealt{2011LRSP....8....6S}, and references therein), flare eruption occurs above the polarity inversion line (PIL) of a solar AR, which separates the adjacent positive polarity and negative polarity of the magnetic field in photosphere. The flaring arcade stretches just over the PIL, and the footpoints of the flaring arcade (as well as the two flare ribbons in chromosphere) are located on the two sides of the PIL. Hence the evolution of the magnetic structures around the PIL are closely related to the flare initiation/eruption processes \citep[e.g.,][]{1993Natur.363..426Z}, and the nonpotential characteristics of magnetic field in a solar AR were often analyzed in the vicinity of the PIL, such as through the magnetic shear in photosphere \citep[e.g.,][]{1996ApJ...456..861W, 2003ApJ...595.1296L} and the twist of field lines in corona \citep[e.g.,][]{2016ApJ...820..103S}. \citet{2014ApJ...788...60J, 2013A&A...555A..77J} and \citet{2012A&A...543A.110A, 2013A&A...549A..66A} extended the classic standard model of solar flares and proposed a three-dimensional (3D) solar flare model. Based on the 3D flare model, they analyzed the strong-to-weak shear transition in post-flare loops \citep{2012A&A...543A.110A}, the upper limit on flare energy \citep{2013A&A...549A..66A}, the reconnection properties of flares \citep{2013A&A...555A..77J}, and the time-evolution of the geometry of photospheric electric currents during the flare eruption \citep{2014ApJ...788...60J}. \citet{1996A&A...308..643D} used the concept of quasi-separatrix layers (QSLs; \citealp{1995JGR...10023443P}) to explore the magnetic configurations of solar flares. They showed that at the locations defined by the QSL method, concentrated electric current build-up can be induced by smooth photospheric motions. As the current density increases, rapid energy release of flares via reconnection can start; but the precise mechanism of this process has yet to be found \citep{1996A&A...308..643D}.

These kinds of studies for investigating the relationship between magnetic structures and flare eruptions require full 3D data of the magnetic field in the solar atmosphere. Currently, only the vector magnetic field at the photosphere level of ARs can be measured with high precision and high resolution, especially by the space-based facilities such as the Hinode satellite \citep{2007SoPh..243....3K} and the Solar Dynamics Observatory (SDO) satellite \citep{2012SoPh..275....3P}. For the magnetic field in corona, one has to rely on the physical modeling and numerical computation to obtain the magnetic field distributions in 3D space, in which the observed photospheric magnetic field data are utilized as the bottom boundary condition \citep[e.g.,][]{1989SSRv...51...11S}. The sophisticated and commonly used physical model for calculating the 3D coronal magnetic field distributions is the nonlinear force-free field (NLFFF) model \citep[e.g.,][]{2012LRSP....9....5W, 2013SoPh..288..481R}, which well represents the physical state of steady corona \citep[e.g.,][]{2005psci.book.....A}.

In NLFFF model for corona, magnetic force dominates and other forces are omitted. The magnetic field vector $\textit{\textbf{B}}$ satisfies $\nabla\times\textit{\textbf{B}}=\alpha\textit{\textbf{B}}$ and $\nabla\alpha\cdot\textit{\textbf{B}}=0$, where the parameter $\alpha$ is a scalar function of spatial position and is called force-free factor \citep{2012LRSP....9....5W, 2013SoPh..288..481R}. The equation $\nabla\times\textit{\textbf{B}}=\alpha\textit{\textbf{B}}$ means that the direction of the electric current density
\begin{equation}\label{equ:j}
  \textit{\textbf{j}}=\frac{1}{4\pi}\nabla\times\textit{\textbf{B}}\ \ (\textrm{in electromagnetic cgs units})
\end{equation}
is parallel ($\alpha>0$) or anti-parallel ($\alpha<0$) to the direction of the magnetic field $\textit{\textbf{B}}$, i.e.,
\begin{equation}\label{equ:j-B}
  \textit{\textbf{j}}=(\alpha/4\pi)\textit{\textbf{B}},
\end{equation}
and hence the corona system is in equilibrium because the Lorentz force
\begin{equation}\label{equ:f}
  \textit{\textbf{f}}=\textit{\textbf{j}}\times\textit{\textbf{B}}
\end{equation}
in this situation is zero (so called force-free). The equation $\nabla\alpha\cdot\textit{\textbf{B}}=0$ means that the value of $\alpha$ is invariant along one individual field line, which is a particular property of the NLFFF and can be utilized for analyzing the magnetic connectivity in corona \citep[e.g.,][]{2014JGRA..119.3286H}.

Since the NLFFF model is built on the steady condition of corona, the dynamic process of flare eruption cannot be simulated by the NLFFF model directly. However, before and after flare eruption, the corona is in quasi-equilibrium states and can be well approximated by the NLFFF model \citep{1989SSRv...51...11S, 2012LRSP....9....5W, 2013SoPh..288..481R}. By investigating the variation of the coronal magnetic fields before and after flare eruption, the changes of the internal coronal magnetic structures concerning the eruption process of flares can be revealed \citep[e.g.,][]{2008ApJ...675.1637S, 2012ApJ...760...17I, 2012ApJ...748...77S, 2014JGRA..119.3286H, 2016ApJ...828...83G}. Compared with the compact or complex solar flare events, the two-ribbon flares have the advantages of adequate spatial scale and purity in morphology, and thus is preferable for this sort of study.

In previous work, \citet{2014JGRA..119.3286H} calculated the time-series of coronal magnetic fields based on the NLFFF model for a notable active region, AR 10930, before and after the X3.4-class two-ribbon flare event happening on 2006 December 13, and investigated the magnetic configuration associated with the flare by analyzing the spatial distribution variations of the electric current density $\textit{\textbf{j}}$ and the force-free factor $\alpha$ before and after the flare eruption. They found that, for the X3.4 flare of AR 10930, a prominent magnetic connectivity (featured by the co-localized strong negative $\alpha$ and strong current density distributions) was formed along the main PIL of the AR before the flare and was totally broken after the flare eruption, and the two branches of the broken magnetic connectivity, combined with the isolated electric current at the magnetic connectivity breaking site (left over from the strong electric current associated with the prominent magnetic connectivity formed before the flare), compose a Z-shaped configuration. In this work, we make a comparative analysis between the magnetic configurations associated with the X3.4 flare event in AR 10930 and the X2.2 two-ribbon flare event happening on 2011 February 15 in another notable active region, AR 11158. We are interested in the comparison on the magnetic configurations of the two flares because the magnetic fields in the corresponding ARs (i.e., AR 10930 and AR 11158) have opposite chiral characteristics in the flare source regions around the main PILs. In short, the magnetic chirality around the main PIL is left-handed for AR 10930 and right-handed for AR 11158 (see the detailed analysis and diagram illustration in Section \ref{sec:chiral}). In literature, the chirality of solar magnetic features were often discussed in the context of chromospheric filaments \citep[e.g.,][]{1998SoPh..182..107M}, magnetic helicity \citep[e.g.,][]{1994ESASP.373...39R, 2005HiA....13...89P, 2014SSRv..186..285P}, and sigmoids of coronal loops \citep[e.g.,][]{1999GeoRL..26..627C, 2007ApJ...671L..81C}. In this work, we want to know what the discrepancies of the flare magnetic configurations are in flare source regions with opposite magnetic chirality.

In Section \ref{sec:events}, we summarize the basic information of the two major two-ribbon solar flare events in AR 10930 and AR 11158. In Section \ref{sec:corona-j}, we display the preflare and postflare photospheric vector magnetograms of the two ARs as well as the corresponding coronal magnetic fields calculated based on the NLFFF model. Then the magnetic nonpotentiality in the two ARs are discussed on the basis of the derived coronal magnetic field data. In Section \ref{sec:chiral}, we explore the chiral characteristics of the magnetic fields in the flare source regions of the two ARs. In Section \ref{sec:confi}, we investigate and compare the magnetic configurations associated with the two flare events. More discussions on the magnetic configurations associated with the two-ribbon flares are given in Sections \ref{sec:discus}. Section \ref{sec:conclu} is the conclusion.

\section{Two major solar flare events} \label{sec:events}

Two major solar flare events are employed for the comparative study on the chirality and magnetic configuration associated with two-ribbon flares: the X3.4-class flare event of AR 10930 on 2006 December 13 and the X2.2-class flare event of AR 11158 on 2011 February 15. Both the two flare events are classic two-ribbon flares and had been studied by many authors in literature (see e.g., \citealt{2007PASJ...59S.779K, 2007ApJ...662L..35Z, 2008ApJ...679.1629G, 2008ApJ...675.1637S, 2012ApJ...760...17I, 2014JGRA..119.3286H} for the flare of AR 10930 and e.g., \citealt{2011ApJ...738..167S, 2012ApJ...748...77S, 2012ApJ...745L..17W, 2012ApJ...745L...4L} for the flare of AR 11158).  The two flares, together with their accompanied CMEs, caused significant disturbances to the solar-terrestrial space weather conditions \citep[e.g.,][]{2009RaSc...44.0A25C, 2009ApJ...704..469M, 2018ApJ...860...26R}.

As introduced in Section \ref{sec:intro}, we chose the two flare events because the magnetic fields in the flare source regions of the two ARs have opposite chirality (i.e., left-handed for the flare in AR 10930 and right-handed for the flare in AR 11158; see Section \ref{sec:chiral} for the detailed chirality analysis), thus they are expected to show different magnetic configurations. On the other hand, as the classic two-ribbon flares, both the flare events have adequate spatial scale and purity in morphology and hence are favourable for exhibiting the physical essence compared with the more compact or more complex flares.

\begin{figure}[htbp]
  \centering
  \includegraphics[scale=0.70]{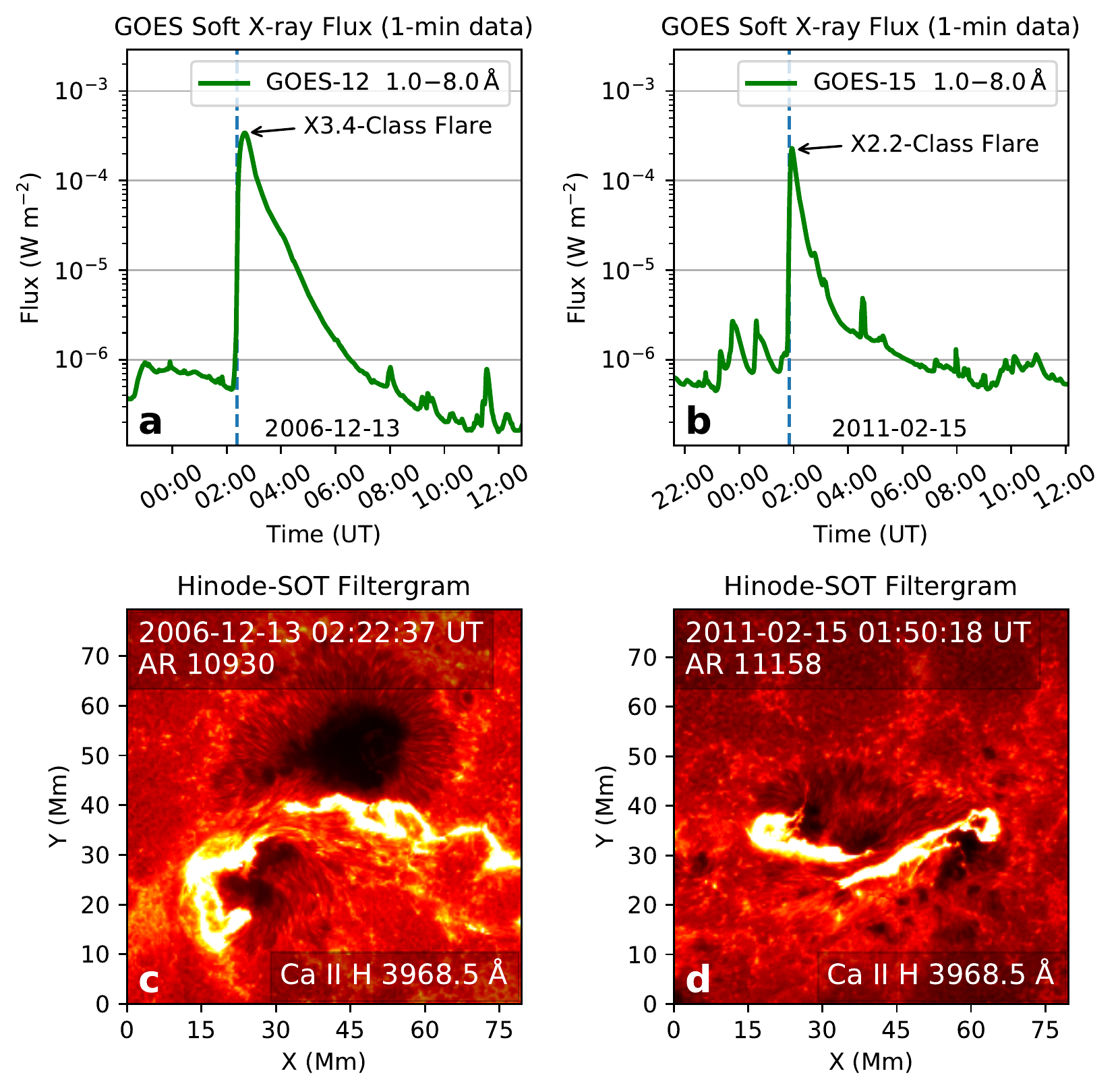}
  \caption{GOES soft X-ray light curves (1-minute averaged data in 1.0--8.0~{\AA} band; top panels) and Hinode-SOT flare-ribbon images (Ca \textsc{ii} H 3968.5~{\AA} filtergrams; bottom panels) of the two major solar flare events in AR 10930 and AR 11158 (see Table \ref{tab:flareinfo}). Left column is for the X3.4-class flare of AR 10930 and right column is for the X2.2-class flare of AR 11158. The soft X-ray flux data of the two flares were acquired by the GOES-12 and GOES-15 satellites, respectively. The vertical dashed lines in the top panels indicate the observation times of the two flare-ribbon images in the context of the soft X-ray light curves. \label{fig:fig1}}
\end{figure}

Table \ref{tab:flareinfo} summarizes the basic information (source active region, class, date, start time, peak time, location, attribute, and magnetic chirality) of the two major flare events in AR 10930 and AR 11158. The soft X-ray light curves (1-minute averaged data in 1.0--8.0~{\AA} band) of the two flares are shown in the top panels of Figure \ref{fig:fig1}. The data of the solar soft X-ray fluxes were acquired by the Geostationary Operational Environmental Satellites (GOES) and provided by the National Centers for Environmental Information (NCEI, formerly the National Geophysical Data Center; see \url{https://www.ngdc.noaa.gov/stp/satellite/goes/index.html}) of the National Oceanic and Atmospheric Administration (NOAA). The bottom panels of Figure \ref{fig:fig1} display the flare-ribbon images of the two flares observed in chromosphere, which were captured by Solar Optical Telescope (SOT; \citealt{2008SoPh..249..167T}) aboard the Hinode satellite \citep{2007SoPh..243....3K} through the Ca \textsc{ii} H spectral line (3968.5~{\AA}).

\begin{table*}[htbp]
  \small
  \begin{threeparttable}
  \caption{Basic information of the two major solar flare events in AR 10930 and AR 11158 \label{tab:flareinfo}}
  \begin{tabular}{cccccccc}
  \hline
  Source & Class\tnote{b} & Date & Start & Peak & Location\tnote{c} & Attribute & Magnetic \\
  active region\tnote{a} &  &  & time\tnote{b} & time\tnote{b} &  &  & chirality\tnote{d} \\
  \hline
  AR 10930 & X3.4 & 2006 December 13 & 02:14 UT & 02:40 UT & S06W22 & two-ribbon; CME & left-handed\\
  AR 11158 & X2.2 & 2011 February 15 & 01:44 UT & 01:56 UT & S20W12 & two-ribbon; CME & right-handed\\
  \hline
  \end{tabular}
  \begin{tablenotes}
  \item[a]{The serial numbers of the solar ARs were assigned and issued by the Space Weather Prediction Center (SWPC) of NOAA (see \url{https://www.swpc.noaa.gov/products/solar-region-summary}).}
  \item[b]{The classes, start times, and peak times of the two flare events were compiled and issued by the SWPC of NOAA (see \url{https://www.swpc.noaa.gov/products/solar-and-geophysical-event-reports}), which were determined based on the 1-minute averaged 1.0--8.0~{\AA} solar soft X-ray flux data (see top panels of Figures \ref{fig:fig1}) acquired by the GOES series of satellites.}
  \item[c]{The flare locations on the solar disk are expressed in degrees (heliographic coordinate system). `S' means south from the solar equator and `W' means west from the central meridian. The coordinate values of the X3.4 flare in AR 10930 were derived from the solar {195~\AA} full-disk image data observed at 2006-12-13 02:36:09 UT by the Extreme-ultraviolet Imaging Telescope (EIT; \citealt{1995SoPh..162..291D}) aboard the Solar and Heliospheric Observatory (SOHO) spacecraft \citep{1995SoPh..162....1D}; and the coordinate values of the X2.2 flare in AR 11158 were derived from the solar {94~\AA} full-disk image data observed at 2011-02-15 01:56:02 UT by the Atmospheric Imaging Assembly (AIA) instrument \citep{2012SoPh..275...17L} aboard the SDO satellite \citep{2012SoPh..275....3P}.}
  \item[d]{See Section \ref{sec:chiral} for the detailed analysis of the magnetic chirality associated with the two flares.}
\end{tablenotes}
\end{threeparttable}
\end{table*}

It can be seen from the flare-ribbon images in Figure \ref{fig:fig1} (bottom panels) that the orientations of the flare-ribbon distributions (i.e., the pointing direction from one ribbon to another ribbon) are different for the two flare events. For the flare of AR 10930, the orientation of the flare-ribbon distribution is from lower-left to upper-right (or right-leaning; see Figure \ref{fig:fig1}c); while for the flare of AR 11158, the orientation of the flare-ribbon distribution is from lower-right to upper-left (or left-leaning; see Figure \ref{fig:fig1}d). The morphology discrepancy of the flare-ribbon distributions implies the different chiral characteristics of the magnetic fields in the source regions of the two flares (see Section \ref{sec:chiral} for the detailed analysis of the magnetic chirality in the flare source regions of the two ARs).

\section{Coronal magnetic field and nonpotentiality in the source ARs of the two flares} \label{sec:corona-j}

To analyze and compare the magnetic structures associated with the two flare events in AR 10930 and AR 11158, for each flare source AR, we selected one photospheric vector magnetogram before the flare and one magnetogram after the flare eruption, and calculated the corresponding coronal magnetic field distributions from the photospheric vector magnetograms based on the NLFFF model. The algorithm of the NLFFF modeling is an improvement to the direct boundary integral equation (DBIE) approach suggested by \citet{2006ApJ...638.1162Y}. (DBIE, in turn, is an advancement to the original BIE method proposed by \citealt{2000SoPh..195...89Y}, see also \citealt{2000SoPh..197..263W, 2001SoPh..201..323W, 2006MNRAS.369..207H, 2011JGRA..116.1101H}.) The implementation of the NLFFF algorithm was described in detail in the papers by \citet{2008JGRA..113.5S90H} and \citet{2011JGRA..116.1101H}. This NLFFF modeling method has been utilized for analyzing the coronal magnetic structures of a variety of solar ARs \citep[e.g.,][]{2008JGRA..113.5S90H, 2011JGRA..116.1101H, 2012agos...30..117H, 2014JGRA..119.3286H, 2013IAUS..294..569L, 2013SoPh..288..507W, 2014ApJ...786...72Y, 2015SCPMA..58.5682W}.

\begin{figure}[htbp]
  \centering
  \includegraphics[scale=0.72]{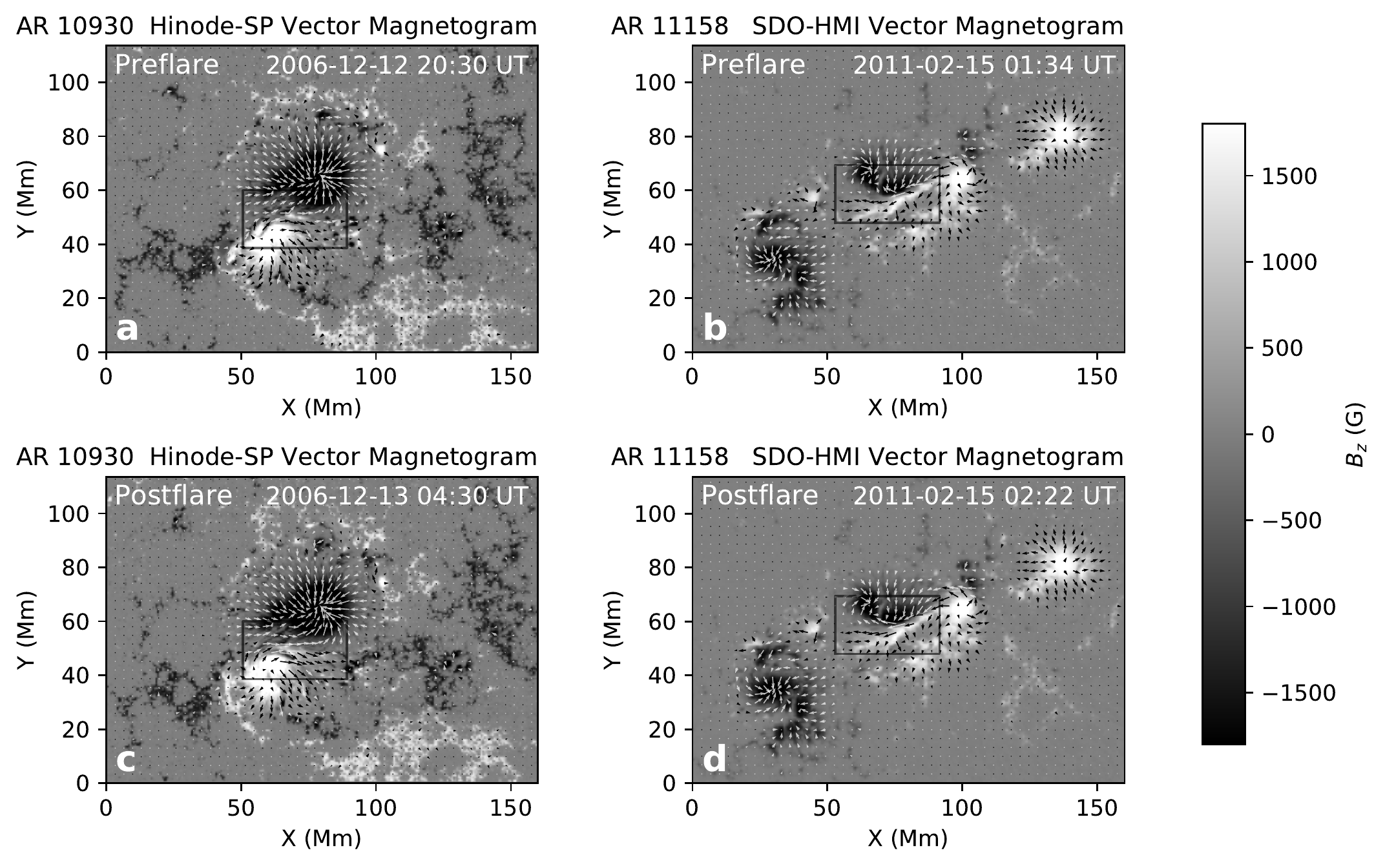}
  \caption{Photospheric vector magnetograms of AR 10930 and AR 11158 before and after the flare eruptions. The left column is for AR 10930 and the right column is for AR 11158. The top row is for the preflare magnetograms and the bottom row is for the postflare magnetograms. The gray scale images display the vertical component ($B_z$) of the magnetograms (white color representing positive polarity and black color representing negative polarity; the gray scale saturating at $\pm 1800$ G). The small arrows overlying the gray scale images represent the transverse component ($\vec{B}_t$) of the vector magnetograms. The field of view and pixel scale of the displayed magnetograms are $160 \times 114$ Mm (corresponding to $224'' \times 159''$) and 0.36 Mm/pixel (corresponding to $0.5''$/pixel) for both the ARs. The black box in each panel indicates the main PIL zone of the corresponding magnetogram. The magnetogram data of AR 10930 were acquired by the Hinode-SP instrument \citep{2013SoPh..283..579L}, and the magnetogram data of AR 11158 were acquired by the SDO-HMI instrument \citep{2012SoPh..275..207S}. The projection effect in the original magnetic field data had been corrected. \label{fig:fig2}}
\end{figure}

Figure \ref{fig:fig2} displays the employed photospheric vector magnetograms of AR 10930 and AR 11158 before and after the flare eruptions. The vertical component of the photospheric vector magnetograms (denoted by $B_z$) is shown by the gray scale images (white color representing positive polarity and black color representing negative polarity), and the transverse component of the vector magnetograms (denoted by $\vec{B}_t$) is visualized by the small arrows overlying the $B_z$ images. The data of the photospheric magnetic fields of AR 10930 and AR 11158 were acquired by the Spectro-Polarimeter (SP) instrument of Hinode \citep{2013SoPh..283..579L} and the Helioseismic and Magnetic Imager (HMI) instrument of SDO \citep{2012SoPh..275..207S}, respectively. For AR 10930, we used the Level 2 data of Hinode-SP \citep{2013SoPh..283..601L} to yield the regular photospheric vector magnetograms, as described in the paper by \citealt{2014JGRA..119.3286H}. For AR 11158, we adopt the hmi.sharp\_cea\_720s photospheric magnetogram data product of the SDO-HMI vector magnetic field pipeline \citep{2014SoPh..289.3549B, 2014SoPh..289.3531C, 2014SoPh..289.3483H}. The preflare magnetograms of the two ARs (top row of Figure \ref{fig:fig2}) are the latest photospheric vector magnetograms obtained by the respective facilities before the start times (see Table \ref{tab:flareinfo}) of the two flares. The postflare magnetograms (bottom row of Figure \ref{fig:fig2}) correspond to the times that the flare-ribbons of the two flares (as shown in the Ca \textsc{ii} H filtergrams observed by Hinode-SOT) started to dissipate. The field of view of the photospheric vector magnetograms employed in this work is $160 \times 114$ Mm (corresponding to $224'' \times 159''$) for both the ARs (see Figure \ref{fig:fig2}). The projection effect in the original magnetic field data had been corrected.

To keep the compatibility between the Hinode-SP data and the SDO-HMI data, we rebinned the SP magnetograms to a similar spatial resolution as that of the HMI data (about $0.5''$/pixel or 0.36 Mm/pixel), as displayed in Figure \ref{fig:fig2}. All the magnetograms were further rebinned to $1''$/pixel resolution (for reducing the burden of computation) to serve as the input data of the NLFFF modeling. The azimuth ambiguity in the photospheric vector magnetic field data were resolved with the nonpotential magnetic field calculation (NPFC) technique \citep{2005ApJ...629L..69G} for the Hinode-SP data, and with the Minimum Energy (ME0) approach \citep{1994SoPh..155..235M, 2009SoPh..260...83L} for the SDO-HMI data as adopted by the HMI team \citep{2014SoPh..289.3483H}. In this work, our analysis is focused on the main PIL zones of the magnetograms (see Figure \ref{fig:fig2} and the descriptions below), where the photospheric magnetic field is relatively strong and the noise level is relatively low, thus the impacts of the different instruments and the azimuth disambiguation methods on the analysis results are minimized (see e.g., \citealt{2017ApJ...851..111S} for a comprehensive investigation on this issue).

From the photospheric vector magnetograms of AR 10930 and AR 11158 shown in Figure \ref{fig:fig2}, it can be seen that the main PILs between the dominant positive polarity and negative polarity of $B_z$ were well developed for both the ARs. The zones of the main PILs in the magnetograms are indicated by black boxes in Figure \ref{fig:fig2}. Along the main PILs, the directions of $\vec{B}_t$ are nearly parallel to the orientations of the PILs, which represents strong magnetic shear. The strong magnetic shear implies strong magnetic nonpotentiality in the two ARs \citep{1996ApJ...456..861W, 2003ApJ...595.1296L}.

\begin{figure}[htbp]
  \centering
  \includegraphics[scale=0.26]{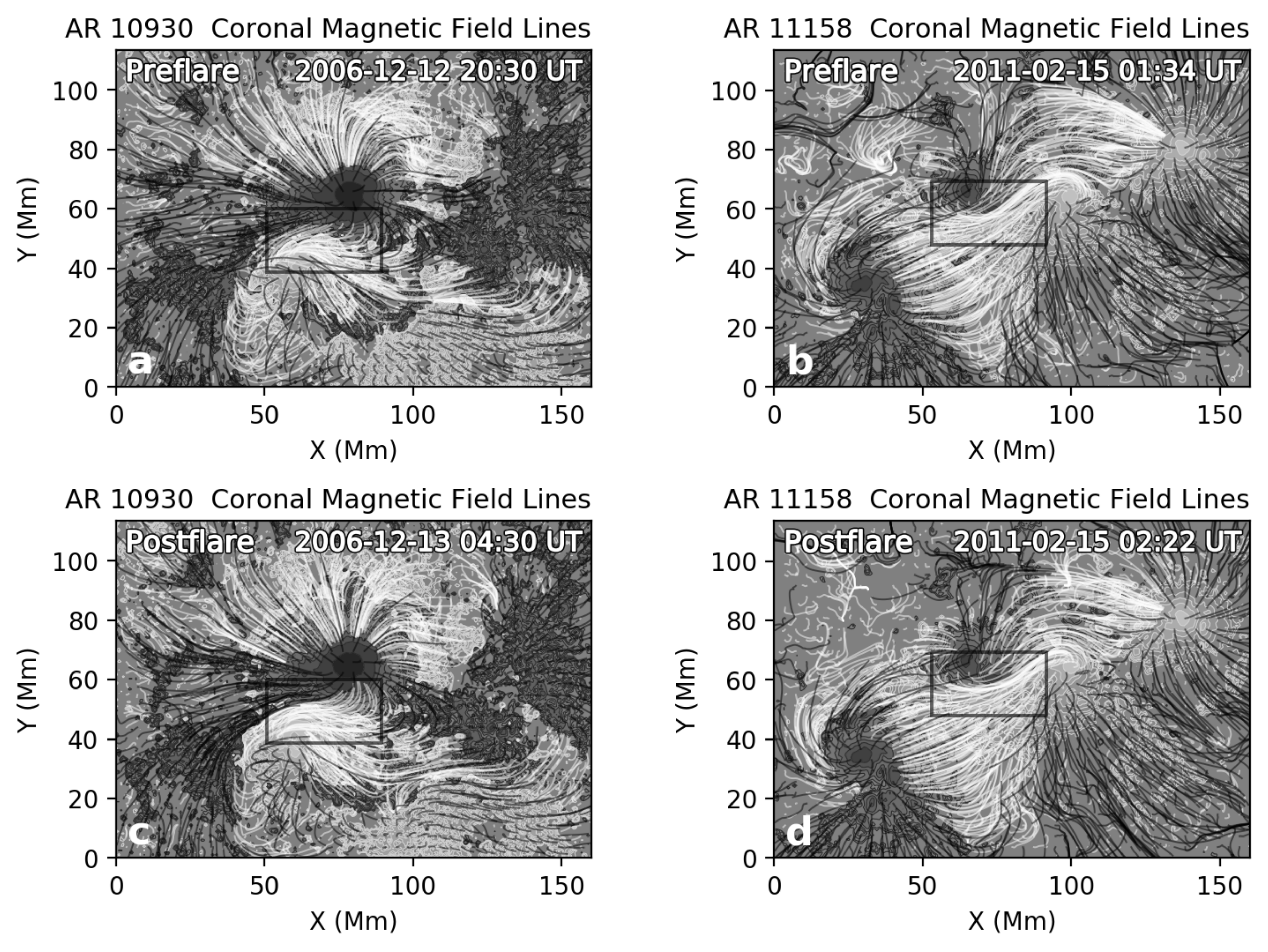}
  \caption{Field line diagrams illustrating the coronal magnetic fields in AR 10930 and AR 11158. The four panels correspond to the four photospheric vector magnetograms shown in Figure \ref{fig:fig2}. The coronal magnetic field data were calculated based on the NLFFF model, in which the photospheric vector magnetogram data were employed as the bottom boundary condition. The $B_z$ component of the photospheric magnetograms is displayed as the background contour plot in each panel. The field lines were traced up to 21 Mm in height. The closed field lines are in white, and the open field lines are in black. The black box in each panel indicates the main PIL zone of the corresponding photospheric magnetogram (same box position as in Figure \ref{fig:fig2}). \label{fig:fig3}}
\end{figure}

The field line diagrams of the coronal magnetic fields (after NLFFF modeling) corresponding to the four photospheric vector magnetograms are illustrated in Figure \ref{fig:fig3}. It can be seen from Figure \ref{fig:fig3} that the magnetic field lines of both ARs show strong twist around the main PILs (indicated by the black boxes in Figure \ref{fig:fig3}) before the flare, and these highly twisted field lines relaxed to some extent after the flare (see Section \ref{sec:discus} for further discussion about the field line relaxation). The strong twist of field lines also indicate strong magnetic nonpotentiality in the two ARs \citep{2016ApJ...820..103S}.

It can also be seen from Figure \ref{fig:fig3} that, the directions of the coronal magnetic fields along the field lines just above the main PILs are consistent with the directions of $\vec{B}_t$ along the main PILs in the photosphere (see Figure \ref{fig:fig2}). In AR 10930, the direction of $\vec{B}_t$ along the main PIL is roughly rightward (see left panels in Figure \ref{fig:fig2}) and the field lines starting from the positive polarity turn right above the main PIL, which leads to the inverse S-shaped twist orientation of the field lines (see left panels in Figure \ref{fig:fig3}). In AR 11158, the direction of $\vec{B}_t$ along the main PIL is roughly leftward (see right panels in Figure \ref{fig:fig2}) and the field lines starting from the positive polarity turn left above the main PIL, which leads to the S-shaped twist orientation of the field lines (see right panels in Figure \ref{fig:fig3}). The different twist orientations of the field lines around the main PILs imply the different chiral characteristics of the associated magnetic fields in the two ARs (see Section \ref{sec:chiral} for the detailed description of the magnetic chirality). In observations, the twist of field lines correspond to the sigmoids of coronal loops as seen in the solar soft X-ray images, which also exhibit the chiral character and the S-shaped or inverse S-shaped pattern \citep[see e.g.,][]{1999GeoRL..26..627C, 2007ApJ...671L..81C}.

\begin{figure}[htbp]
  \centering
  \includegraphics[scale=0.72]{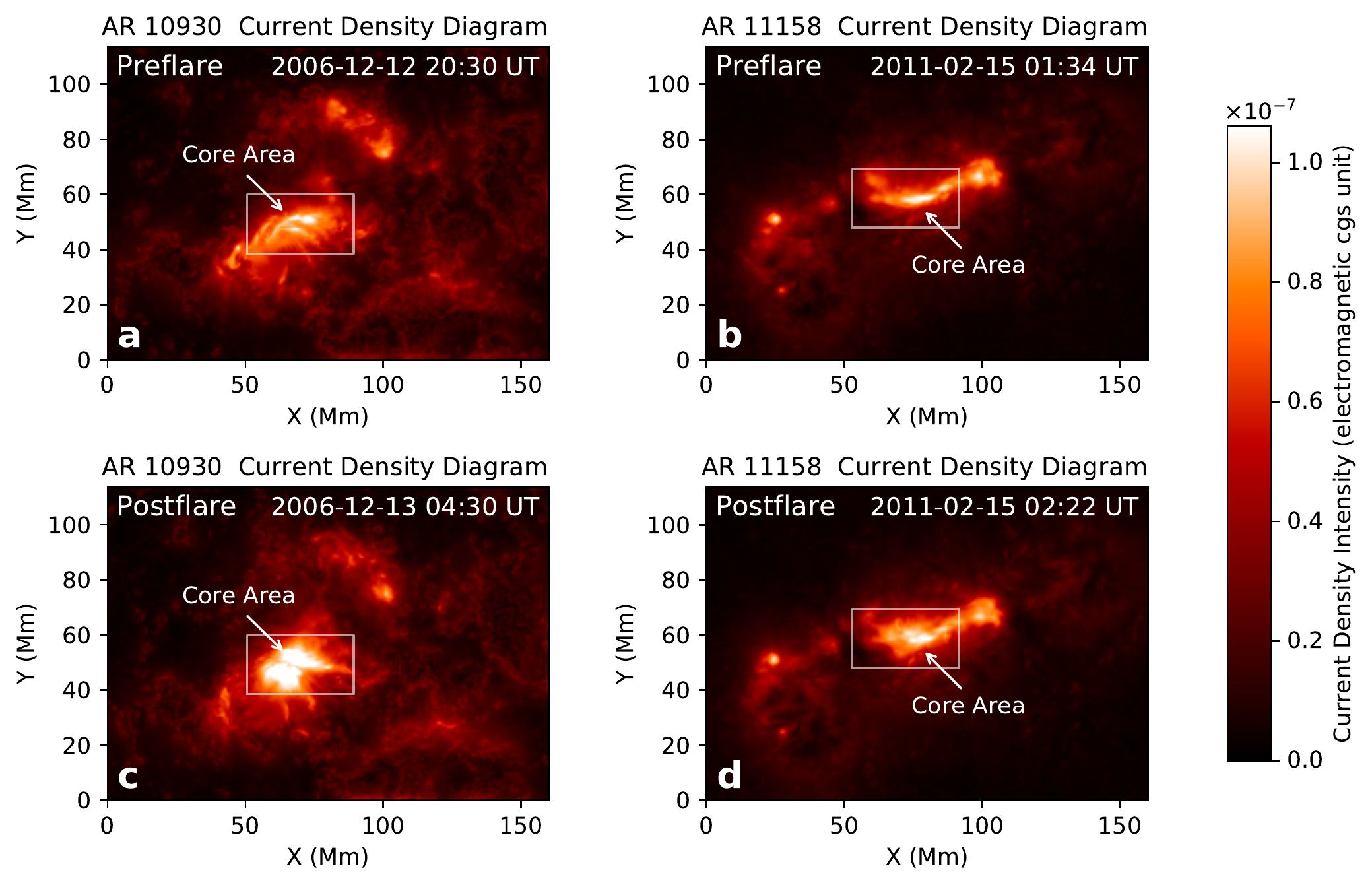}
  \caption{The projected current density distributions in AR 10930 and AR 11158. The four panels correspond to the four coronal magnetic field diagrams shown in Figure \ref{fig:fig3}. The intensity values were obtained by vertically averaging $|\textit{\textbf{j}}|$ in a modeling volume of 21 Mm in height. The white box in each panel indicates the main PIL zone of the corresponding photospheric magnetogram (same box position as in Figure \ref{fig:fig2}). The local area with the highest current density values in each panel is indicated by a white arrow and labeled with `Core Area'. \label{fig:fig4}}
\end{figure}

To exhibit the nonpotential characteristics of the two ARs more directly, we calculated the values of the electric current density $\textit{\textbf{j}}$ from the coronal magnetic field data by using Equation (\ref{equ:j}). The spatial distributions of the derived current density in the two ARs are illustrated in Figure \ref{fig:fig4}, where the four panels correspond to the four coronal magnetic field diagrams shown in Figure \ref{fig:fig3} (and also the four photospheric magnetograms shown in Figure \ref{fig:fig2}). It can be seen from Figure \ref{fig:fig4} that the electric current in the two ARs is distributed mostly in the zones around the main PILs (indicated by the white boxes in Figure \ref{fig:fig4}), which is consistent with the strong twist of field lines around the main PILs shown in Figure \ref{fig:fig3} and the strong magnetic shear along the main PILs shown in Figure \ref{fig:fig2}. In particular, there exists an area possessing the highest current density values (indicated by white arrows and labeled with `Core Area' in Figure \ref{fig:fig4}) in each of the ARs. By comparing the current density images with the flare-ribbon images shown in Figure \ref{fig:fig1}, it can be seen that the core areas of current density coincide with the locations of flare onset in-between the flare ribbons. This result indicates that both the two flares initiated in the local regions with strongest current density, which represent the areas with most active level and strongest nonpotentiality in the two ARs.

Figure \ref{fig:fig4} also shows that the patterns of the current density distributions in the core areas of the two ARs (indicated by the white arrows in Figure \ref{fig:fig4}) present apparent changes before (see the top panels of Figure \ref{fig:fig4}) and after (see the bottom panels of Figure \ref{fig:fig4}) the flares, which implies the variations of the internal coronal magnetic structures before and after the flare eruptions. We will investigate and compare the magnetic configurations associated with the two flare events in Section \ref{sec:confi}.

\section{Chirality of the magnetic fields in the flare source regions} \label{sec:chiral}

The photospheric vector magnetograms of AR 10930 and AR 11158 shown in Figure \ref{fig:fig2} exhibit that, for both the ARs, the positive polarity of $B_z$ is roughly below the main PIL and the negative polarity is above the main PIL (see the main PIL zones enclosed by the black boxes in Figure \ref{fig:fig2}). Yet the direction of the transverse magnetic field vectors $\vec{B}_t$ (illustrated by the small arrows in Figure \ref{fig:fig2}) along the main PILs is opposite for the two ARs, that is, $\vec{B}_t$ is roughly rightward for AR 10930 and leftward for AR 11158. This discrepancy represents the opposite chiral characteristics of the magnetic fields in the flare source regions (the main PIL zones) of the two ARs. Specifically, the left-handed or right-handed chirality is defined by the morphological relation between the direction of $\vec{B}_t$ and the polarity orientation of $B_z$ following the left-hand or right-hand rule, in which the thumb points in the direction of $\vec{B}_t$ and the other four fingers point from the positive polarity to the negative polarity of $B_z$ with the palm facing the PIL. Thus, the magnetic chirality in the main PIL zones of the two ARs is left-handed for AR 10930 and right-handed for AR 11158. Note that this definition of the chirality for the photospheric vector magnetic fields is consistent with the chirality definition in literature for the chromospheric filaments \citep[e.g.,][]{1998SoPh..182..107M, 2017ApJ...835...94O} and the magnetic helicity \citep[e.g.,][]{1994ESASP.373...39R, 2005HiA....13...89P, 2014SSRv..186..285P}.

\begin{figure}[htbp]
  \centering
  \includegraphics[scale=0.72]{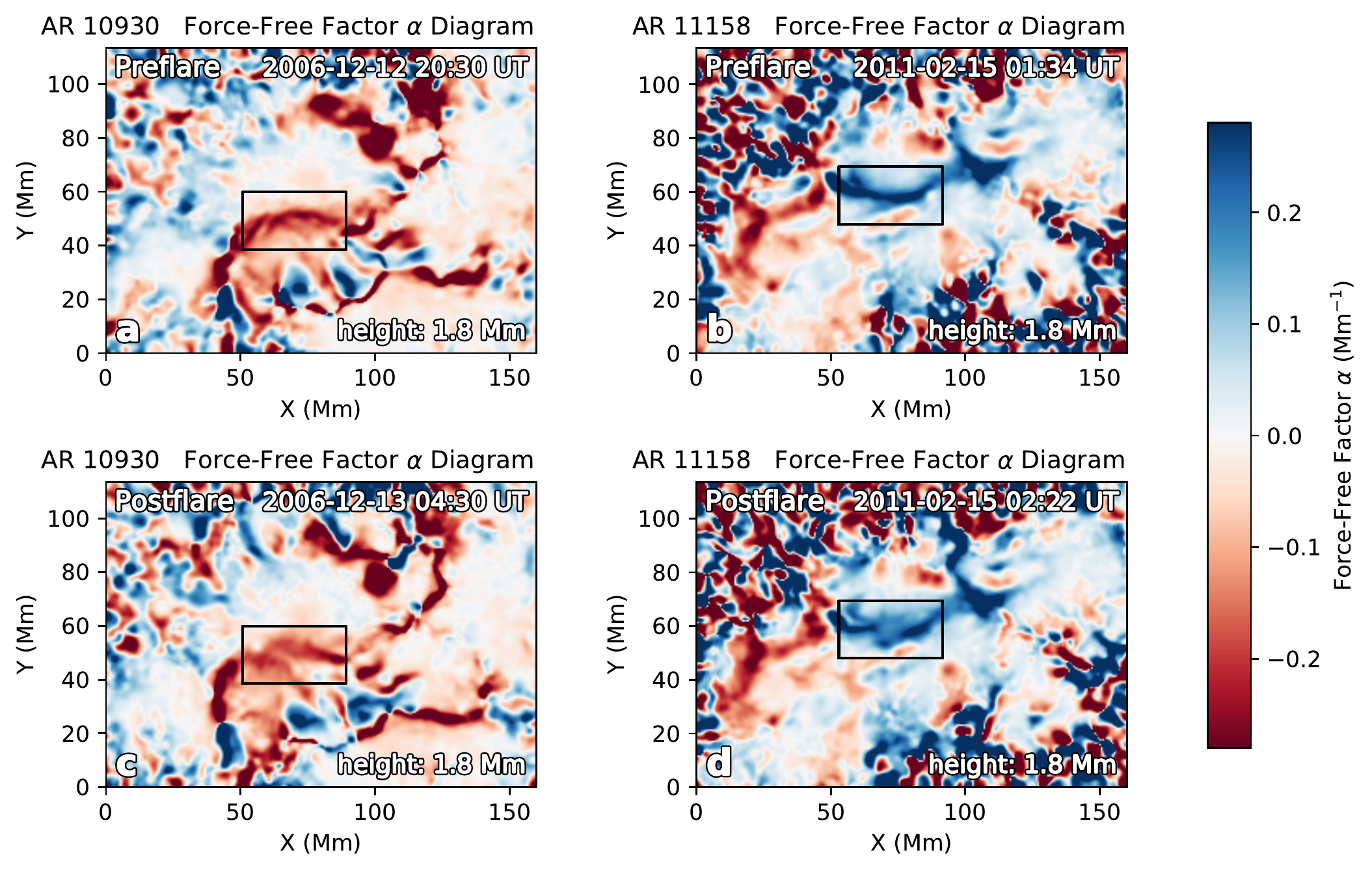}
  \caption{Distributions of force-free factor $\alpha$ at height of 1.8 Mm in AR 10930 and AR 11158. The four panels correspond to the four coronal magnetic field diagrams shown in Figure \ref{fig:fig3}. The negative and positive values of $\alpha$ are rendered in red and blue colors, respectively. The black box in each panel indicates the main PIL zone of the corresponding photospheric magnetogram (same box position as in Figure \ref{fig:fig2}). The $\alpha$ distributions in the marginal areas are noisy owing to the very weak magnetic field there. \label{fig:fig5}}
\end{figure}

The chiral characteristics of the magnetic fields in solar ARs can also be quantitatively reflected by the signs of force-free factor $\alpha$ of the coronal magnetic fields. Considering the equation of the force-free field, $\nabla\times\textit{\textbf{B}}=\alpha\textit{\textbf{B}}$, the values of $\alpha$ can be computed from the coronal magnetic field data by using equation \citep{2011JGRA..116.1101H, 2014JGRA..119.3286H}:
\begin{equation}\label{equ:alpha}
  \alpha=\frac{(\nabla \times \textit{\textbf{B}}) \cdot \textit{\textbf{B}}}{B^2}.
\end{equation}
The order of magnitude of $\alpha$ is about 0.1 Mm$^{-1}$ in the lower layers of corona and decreases rapidly with height (see e.g., \citealt{2014JGRA..119.3286H}). Figure \ref{fig:fig5} illustrates the distributions of force-free factor $\alpha$ in the two ARs at height of 1.8 Mm, where the four panels correspond to the four coronal magnetic field diagrams shown in Figure \ref{fig:fig3} (and also the four photospheric magnetograms shown in Figure \ref{fig:fig2} and the four current density diagrams shown in Figure \ref{fig:fig4}). It can be seen from Figure \ref{fig:fig5} that, along the main PILs of the two ARs (enclosed by the black boxes in Figure \ref{fig:fig5}), $\alpha<0$ for AR 10930 (see the left panels of Figure \ref{fig:fig5}; the negative $\alpha$ being rendered in red color) and $\alpha>0$ for AR 11158 (see the right panels of Figure \ref{fig:fig5}; the positive $\alpha$ being rendered in blue color), which quantitatively reflects the left-handed chirality around the main PIL of AR 10930 and the right-handed chirality of AR 11158.

The correspondence between the signs of force-free factor $\alpha$ and the chirality of the magnetic fields can be understood as follows: In the photospheric vector magnetograms of the two ARs (see Figure \ref{fig:fig2}), the positive polarity of $B_z$ is below the main PILs and the negative polarity is above the main PILs. To be compatible with the polarity distribution of $B_z$, the direction of the electric current density along the main PILs should be leftward for both the ARs according to the Ampere's rule in electromagnetism. In AR 10930 with left-handed chirality, the direction of the magnetic field along the main PIL is rightward (consistent with the direction of $\vec{B}_t$ shown in Figure \ref{fig:fig2}), which is opposite to the leftward direction of the electric current density, and hence $\alpha < 0 $ (see Equation (\ref{equ:j-B})); while in AR 11158 with right-handed chirality, the direction of the magnetic field along the main PIL is leftward, which is the same as the direction of the electric current density, and hence $\alpha > 0 $.

The left-handed or right-handed magnetic chirality around the main PILs of the two ARs can also be reflected by the twist orientation of the coronal magnetic field lines (inverse S-shaped or S-shaped, see the diagram illustration in Figure \ref{fig:fig3} and description in Section \ref{sec:corona-j}) and the orientation of the flare-ribbon distribution (right-leaning or left-leaning, see the flare-ribbon images in Figures \ref{fig:fig1}c and \ref{fig:fig1}d and description in Section \ref{sec:events}), which are all consistent with the directions of $\vec{B}_t$ along the main PILs in the photosphere (see the photospheric vector magnetograms in Figure \ref{fig:fig2}). Table \ref{tab:chirality} summarizes the correspondence between the various features representing magnetic chirality, which include direction of $\vec{B}_t$ in the photosphere, sign of force-free factor $\alpha$, twist orientation of the coronal magnetic field lines, and orientation of the flare-ribbon distribution.

\begin{table*}[htbp]
  \small
  \caption{Correspondence between the various features representing magnetic chirality \label{tab:chirality}}
  \begin{tabular}{lccc}
  \hline
  Feature around the main PIL & & \multicolumn{2}{c}{Magnetic chirality} \\
  \cline{3-4}
   & & Left-handed & Right-handed  \\
   & & (AR 10930) & (AR 11158) \\
  \hline
  Direction of $\vec{B}_t$ in the photosphere (see Figure \ref{fig:fig2}) & & rightward ($\rightarrow$)  & leftward ($\leftarrow$)   \\
  Sign of force-free factor $\alpha$ (see Figure \ref{fig:fig5}) & & negative ($\alpha < 0 $) & positive ($\alpha > 0 $) \\
  Twist orientation of the coronal magnetic field lines (see Figure \ref{fig:fig3}) & &  inverse S-shaped  &
  S-shaped  \\
  Orientation of the flare-ribbon distribution (see Figures \ref{fig:fig1}c and \ref{fig:fig1}d)   & &  right-leaning ($\ \diagup\ $)  &
  left-leaning ($\ \diagdown\ $)  \\
  \hline
  \end{tabular}
\end{table*}

\section{Magnetic configurations associated with the two flares} \label{sec:confi}

The spatial distribution diagrams of the force-free factor $\alpha$ shown in Figure \ref{fig:fig5} not only indicate the chiral characteristics of the magnetic fields in the two ARs by its sign, but also exhibit the magnetic connectivity properties in the coronal magnetic fields of the two ARs. Note that in the force-free magnetic field, $\alpha$ is a constant along each field line (see the description in Section \ref{sec:intro}). It can be seen in Figure \ref{fig:fig5} that, for both the ARs, the patterns of the $\alpha$ distributions in the main PIL zones of the ARs (see the regions enclosed by the black boxes in Figure \ref{fig:fig5}) have apparent changes before (top panels of Figure \ref{fig:fig5}) and after (bottom panels of Figure \ref{fig:fig5}) the flares, which from another point of view (in addition to the distribution variations of the electric current density $\textit{\textbf{j}}$ shown in Figure \ref{fig:fig4}) exhibits the variations of the internal coronal magnetic structures before and after the flare eruptions. Note that in corona the electric current density $\textit{\textbf{j}}$, the force-free factor $\alpha$, and the magnetic field $\textit{\textbf{B}}$ are connected by Equation (\ref{equ:j-B}). Relative to the fundamental magnetic field data shown in Figures \ref{fig:fig2} and \ref{fig:fig3}, the derived differential quantities $\alpha$ and $\textit{\textbf{j}}$ shown in Figures \ref{fig:fig4} and \ref{fig:fig5} can reveal the variations of the internal magnetic structures more explicitly. In this work, we investigate and compare the magnetic configurations associated with the two flare events in AR 10930 and AR 11158 through the spatial distributions of $\alpha$ and $\textit{\textbf{j}}$.

\begin{figure}[htbp]
  \centering
  \includegraphics[scale=0.53]{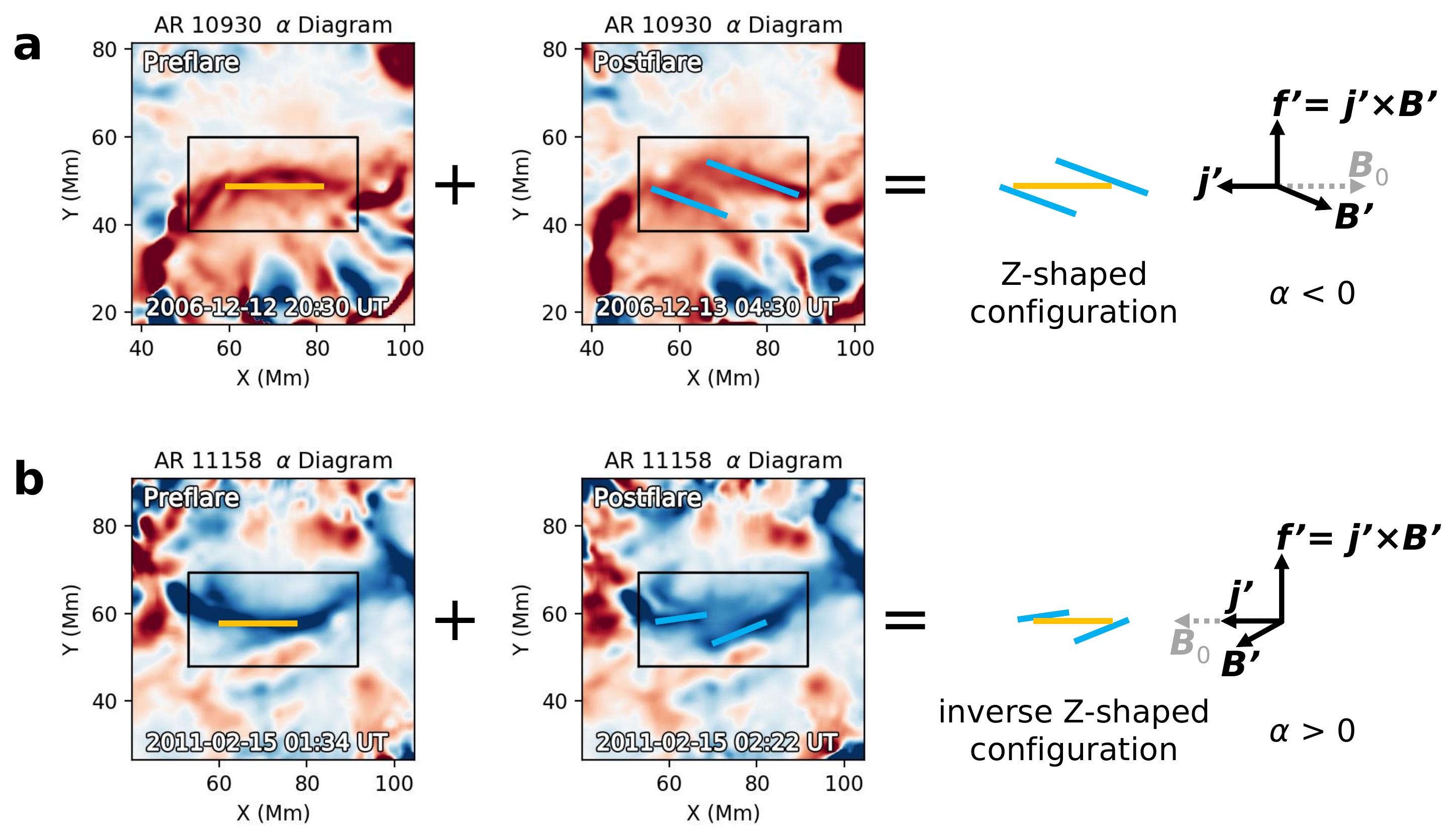}
  \caption{Diagrams illustrating (a) the Z-shaped magnetic configuration of the X3.4 flare event of AR 10930 and (b) the inverse Z-shaped magnetic configuration of the X2.2 flare event of AR 11158. The subplots on the left of the `=' symbols show the $\alpha$ distribution maps in the central areas of the two ARs before and after the flare eruptions. The black boxes in the $\alpha$ diagrams indicate the main PIL zones, which are the same as in Figure \ref{fig:fig5}. The bars in golden color in the preflare $\alpha$ diagrams represent the prominent magnetic connectivities as well as the associated strong electric currents (characterized by the highest current density values in the core areas shown in Figure \ref{fig:fig4}) formed before the flares; and the bars in light-blue color in the postflare $\alpha$ diagrams indicate the two branches of the broken magnetic connectivity. The sketches on the right of the `=' symbols illustrate the Z-shaped (for AR 10930) or the inverse Z-shaped (for AR 11158) configuration composed by the two light-blue bars (representing the two branches of the broken magnetic connectivity) and the golden bar (representing the prominent magnetic connectivity or the associated strong electric current before the flare). The golden bar in the Z-shaped or inverse Z-shaped diagrams also represents the isolated electric current at the magnetic connectivity breaking site left over from the strong electric current formed before the flare (associated with the preflare prominent magnetic connectivity). The rightmost schematic diagrams illustrate the upward Lorentz force $\textit{\textbf{f}}\,'=\textit{\textbf{j}}\,' \times \textit{\textbf{B}}\,'$ acting on the isolated electric current (represented by the current density vector $\textit{\textbf{j}}\,'$) at the magnetic connectivity breaking site. $\textit{\textbf{B}}_0$ (in gray color) shows the direction of the force-free magnetic field before the flare, which is anti-parallel to the direction of $\textit{\textbf{j}}\,'$ for AR 10930 and parallel to the direction of $\textit{\textbf{j}}\,'$ for AR 11158 (note that $\textit{\textbf{j}}\,'$ inherits the direction of the preflare electric current density). $\textit{\textbf{B}}\,'$ is the deflected background magnetic field owing to the breaking of magnetic connectivity, which is roughly aligned with the two branches of broken magnetic connectivity (see \ref{sec:appendix} for diagram illustration). It is this magnetic field deflection from $\textit{\textbf{B}}_0$ to $\textit{\textbf{B}}\,'$ that introduces the local Lorentz force $\textit{\textbf{f}}\,'$ on the isolated electric current and causes the flare initial eruption. \label{fig:fig6}}
\end{figure}

The enlarged $\alpha$ distribution maps in the central areas of the two ARs before and after the flare eruptions are shown in the left part of Figure \ref{fig:fig6} (top row for AR 10930 and bottom row for AR 11158). From the $\alpha$ distribution diagrams before the flares (leftmost column in Figure \ref{fig:fig6}) it can be seen that, for both the ARs, there exists a strip of concentrated strong $\alpha$ distribution along the main PIL (see the regions enclosed by the black boxes), which indicates that a magnetic connectivity was formed along the main PIL before the flare. By comparing with the electric current density distribution diagrams before the flares (see the top panels of Figure \ref{fig:fig4}), it can be found that, for both the ARs, the magnetic connectivity along the main PIL is co-localized with the core area featured by the highest current density values (indicated by the white arrows in Figure \ref{fig:fig4}). We refer to these magnetic connectivity accompanied with the strong electric current as the prominent magnetic connectivity. In Figure \ref{fig:fig6}, the prominent magnetic connectivity as well as the accompanied strong electric current in the two ARs are indicated by golden bars. Note that, because the strengths of the force-free factor $\alpha$ and the current density $\textit{\textbf{j}}$ decrease rapidly with height (see e.g., \citealt{2014JGRA..119.3286H}), the prominent magnetic connectivity characterized by both strong $\alpha$ and strong current density mainly distributes at low altitude.

From the second column of Figure \ref{fig:fig6} it can be seen that, after the flare eruption the prominent magnetic connectivity was totally broken and manifests as two separated branches of strong $\alpha$ distributions (indicated by the two light-blue bars in Figure \ref{fig:fig6}) around the main PIL. This is the same for the two ARs. However, the relative position of the two branches of broken magnetic connectivity is different in the two ARs. For AR 10930 it is lower-left to upper-right, and for AR 11158 it is upper-left to lower-right. As a result, the two branches of the broken magnetic connectivity (represented by the two light-blue bars) combined with the prominent magnetic connectivity or the associated strong electric current formed before the flare (represented by the golden bar) compose a Z-shaped configuration in AR 10930 and an inverse Z-shaped configuration in AR 11158, which is sketched on the right of the `=' symbols in Figure \ref{fig:fig6}. As shown in Figure \ref{fig:fig6}, the Z-shaped magnetic configuration associated with the flare event in AR 10930 corresponds to the left-handed magnetic chirality (negative $\alpha$) in the flare source region around the main PIL of AR 10930, and the inverse Z-shaped magnetic configuration associated with the flare event in AR 11158 corresponds to the right-handed magnetic chirality (positive $\alpha$) of AR 11158. In Section \ref{sec:discus}, we give interpretation and further discussions on the correspondence relationship between the magnetic configuration and the magnetic chirality.

\section{Discussion} \label{sec:discus}

We suggest that the correspondence between the magnetic configurations (Z-shaped or inverse Z-shaped) associated with the flares and the magnetic chirality (left-handed or right-handed) in the flare source regions is concerned with the direction of the local Lorentz force that drives the flare initial eruption, which is sketched in the rightmost column of Figure \ref{fig:fig6} and elucidated below.

In our scenario of flare initiation \citep{2014JGRA..119.3286H}, the breaking of the prominent magnetic connectivity (as a sudden change of topological structure of the coronal magnetic field) is driven by the evolution of the photospheric magnetic field which acts as the bottom boundary condition of the corona system. At the breaking site of the prominent magnetic connectivity, the direction of the background magnetic field (denoted by $\textit{\textbf{B}}\,'$ in the rightmost schematic diagrams of Figure \ref{fig:fig6}) is deflected from the direction of the preflare magnetic field (denoted by $\textit{\textbf{B}}_0$ in Figure \ref{fig:fig6}) and is now roughly aligned with the two branches of broken magnetic connectivity (see \ref{sec:appendix} for diagram illustration of the magnetic field deflection at the magnetic connectivity breaking site before and after the two flares). The strong electric current associated with the prominent magnetic connectivity formed before the flare is no longer consistent with the new background magnetic field $\textit{\textbf{B}}'$ and hence is isolated. The golden bar in the Z-shaped or inverse Z-shaped diagrams shown in the right part of Figure \ref{fig:fig6} also represent this isolated electric current at the magnetic connectivity breaking site.

As demonstrated in the rightmost schematic diagrams of Figure \ref{fig:fig6}, there is an angle between the deflected background magnetic field $\textit{\textbf{B}}\,'$ and the isolated electric current (represented by the current density vector $\textit{\textbf{j}}\,'$ in the diagrams) at the magnetic connectivity breaking site. Then the local corona system is deviated from the force-free condition and becomes unstable, and the deflected background magnetic field $\textit{\textbf{B}}\,'$ exerts a local Lorentz force $\textit{\textbf{f}}\,'$ on the electric current $\textit{\textbf{j}}\,'$:
\begin{equation}\label{equ:f-p}
  \textit{\textbf{f}}\,'=\textit{\textbf{j}}\,'\times\textit{\textbf{B}}\,'.
\end{equation}
Since $\textit{\textbf{j}}\,'$ represents the isolated electric current (at the magnetic connectivity breaking site) left over from the strong electric current formed before the flare, it inherits the direction of the preflare electric current density (parallel or anti-parallel to $\textit{\textbf{B}}_0$). According to the descriptions in Section \ref{sec:chiral}, the direction of $\textit{\textbf{B}}_0$ is opposite to the direction of electric current density in AR 10930 owing to the negative $\alpha$ or left-handed chirality of the AR, and is the same as the direction of electric current density in AR 11158 owing to the positive $\alpha$ or right-handed chirality of the AR. As illustrated in the rightmost diagrams of Figure \ref{fig:fig6}, the direction of $\textit{\textbf{j}}\,'$ is leftward for both the ARs, while the direction of $\textit{\textbf{B}}_0$ is rightward for AR 10930 and leftward for AR 11158. The diagrams in the rightmost column of Figure \ref{fig:fig6} also exhibit that, the deflection of $\textit{\textbf{B}}\,'$ relative to $\textit{\textbf{B}}_0$ is clockwise for AR 10930 owing to the Z-shaped magnetic configuration in the AR, and is anticlockwise for AR 11158 owing to the inverse Z-shaped magnetic configuration in the AR. As a result, the direction of the Lorentz force $\textit{\textbf{f}}\,'$ acting on the isolated electric current $\textit{\textbf{j}}\,'$ is upward for both the ARs (see the rightmost diagrams of Figure \ref{fig:fig6}), which leads to the ejection of the isolated electric current and associated plasmas. Along with the ejection of the electric current, the free magnetic energy associated with the ejected current is released. Then the flare eruption is initiated. Owing to the loss of the ejected electric current and the depletion of the free magnetic energy associated with the ejected current, the field lines surrounding the main PIL show some extent of relaxation after flare eruption, as illustrated in Figure \ref{fig:fig3} and described in Section \ref{sec:corona-j}.

\begin{figure}[htbp]
  \centering
  \includegraphics[scale=0.53]{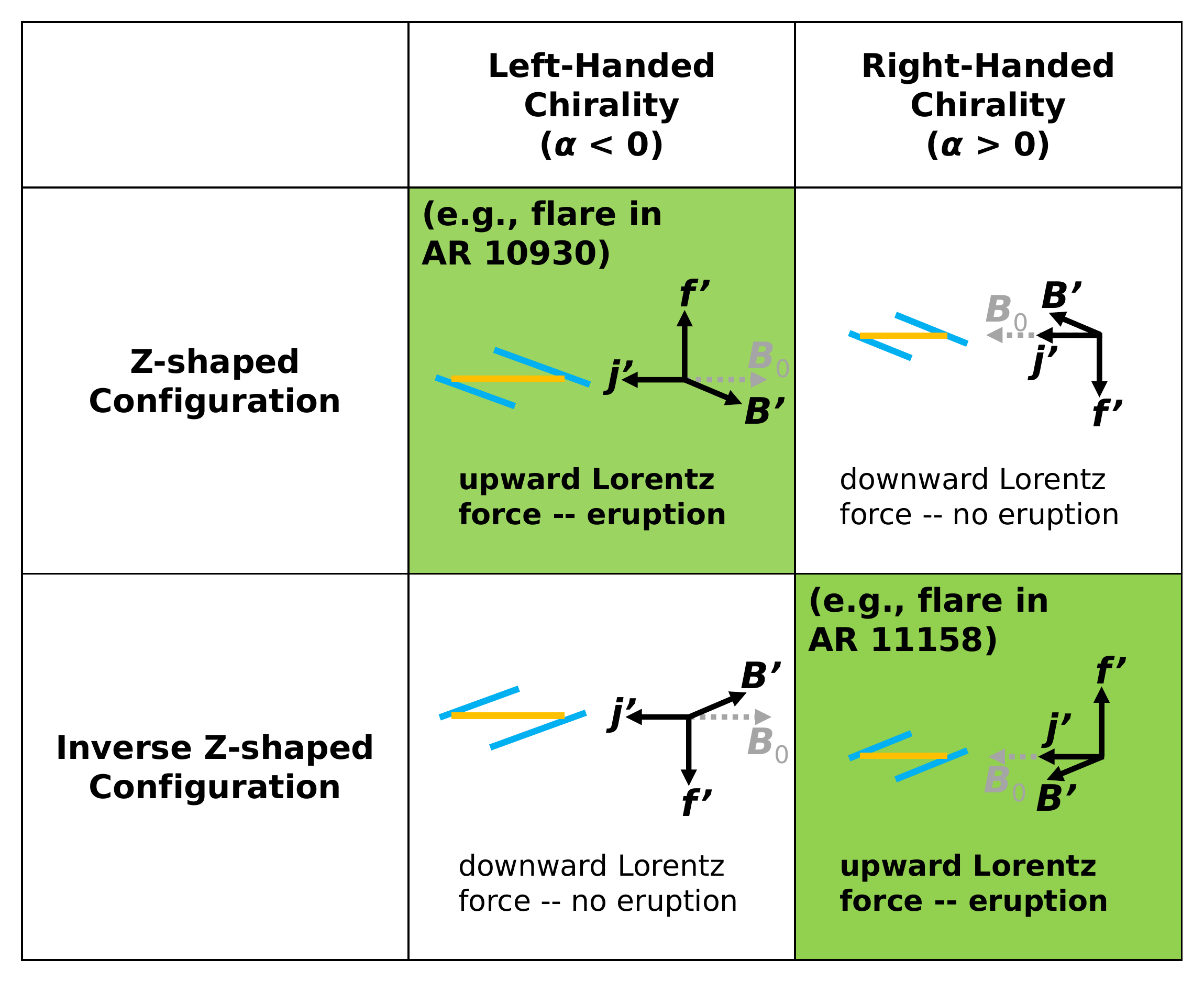}
  \caption{Four possible situations of combination between the magnetic configuration of two-ribbon flares (Z-shaped or inverse Z-shaped) and the magnetic chirality in flare source regions (left-handed or right-handed). The two situations highlighted with green color have upward local Lorentz force $\textit{\textbf{f}}\,'$, and hence can lead to the initial eruption of flares. The other two situations have downward local Lorentz force and cannot lead to eruption. (See Figure \ref{fig:fig6} for the meanings of the symbols and the main text in Section \ref{sec:discus} for the detailed explanation.) \label{fig:fig7}}
\end{figure}

The above discussion shows that, to drive the flare initial eruption, the upward direction of the local Lorentz force acting on the isolated electric current at the magnetic connectivity breaking site requires the correspondence relationship between the magnetic configurations associated with the flares and the magnetic chirality in the flare source regions, that is, the Z-shaped configuration for the left-handed chirality ($\alpha<0$) (e.g., the flare in AR 10930) and the inverse Z-shaped configuration for the right-handed chirality ($\alpha>0$) (e.g., the flare in AR 11158). We speculate that two-ribbon flares can be generally classified to these two magnetic configurations (Z-shaped or inverse Z-shaped) by the chiral characteristics (left-handed or right-handed) of magnetic fields in the flare source regions of ARs. The opposite combinations, i.e., a Z-shaped configuration with a right-handed chirality ($\alpha>0$) or an inverse Z-shaped configuration with a left-handed chirality ($\alpha<0$), are not likely for the two-ribbon flares, because in these cases the local Lorentz force acting on the isolated electric current at the magnetic connectivity breaking site would be downward and cannot lead to eruption. In Figure \ref{fig:fig7}, we summarize the four possible situations of combination between the magnetic configuration and the magnetic chirality.

\section{Conclusion} \label{sec:conclu}

In this paper, we made a comparative analysis on the chiral characteristics and the magnetic configurations associated with two X-class two-ribbon solar flares happening in AR 10930 and AR 11158 (see Table \ref{tab:flareinfo}). The photospheric vector magnetograms of the two ARs (see Figure \ref{fig:fig2}) were observed by the Hinode and the SDO satellites, respectively. The corresponding coronal magnetic fields (see Figure \ref{fig:fig3}) were obtained by the numerical computation based on the NLFFF model. We employed two quantitative measures, the electric current density $\textit{\textbf{j}}$ (see Equation (\ref{equ:j}) and Figure \ref{fig:fig4}) and the force-free factor $\alpha$ (see Equation (\ref{equ:alpha}) and Figure \ref{fig:fig5}), to reflect the internal structural properties of the coronal magnetic fields in the two ARs before and after the flares.

The analysis by the spatial distributions of the electric current density \textit{\textbf{j}} showed that, the electric current in the two ARs was distributed mostly around the main PILs (see Figure \ref{fig:fig4}) where the flares happened. There exists an area possessing the highest current density values in each of the ARs (labeled with `Core Area' in Figure \ref{fig:fig4}), which indicates the local region with most active level and strongest nonpotentiality in the two ARs and hence implies the location of the flare initial eruption. This result of concentrated electric current build-up in flare source region is consistent with the conclusion by \citet{1996A&A...308..643D} via QSL method as introduced in Section \ref{sec:intro}.

The analysis on the chiral characteristics of the magnetic fields in the flare source regions around the main PILs showed that, although the photospheric magnetic fields of the two ARs are complex, the chirality of the magnetic fields around the main PILs in the photosphere (indicated by the morphological relation between the direction of $\vec{B}_t$ and the polarity orientation of $B_z$; see Figure \ref{fig:fig2} and the description in Section \ref{sec:chiral}) and in the corona (indicated by the sign of force-free factor $\alpha$; see Figure \ref{fig:fig5} and the description in Section \ref{sec:chiral}) is coherent for both the ARs, that is, left-handed ($\alpha<0$) for AR 10930 and right-handed ($\alpha>0$) for AR 11158. The magnetic chirality in the flare source regions can also be reflected by the twist orientation of the coronal magnetic field lines (see Figure \ref{fig:fig3}) and the orientation of the flare-ribbon distribution (see Figures \ref{fig:fig1}c and \ref{fig:fig1}d). The correspondence relationship between the various features representing magnetic chirality is summarized in Table \ref{tab:chirality}.

The analysis on the magnetic configurations associated with the flares shows that, for both the flare events, a prominent magnetic connectivity (featured by co-localized strong $\alpha$ and strong electric current density distributions) was formed along the main PIL before flare and was totally broken after flare eruption. Yet the two branches of broken magnetic connectivity, combined with the prominent magnetic connectivity (or the associated strong electric current) before the flare, compose the opposite magnetic configurations for the two flares owing to the opposite magnetic chirality in the flare source regions of the two ARs. That is, Z-shaped configuration for the flare in AR 10930 with left-handed chirality (negative $\alpha$) and inverse Z-shaped configuration for the flare in AR 11158 with right-handed chirality (positive $\alpha$), as illustrated in Figure \ref{fig:fig6}. We suggest that it is the upward local Lorentz force acting on the isolated electric current at the magnetic connectivity breaking site (left over from the strong electric current formed before the flare) that drives the flare initial eruption, and the requirement of the upward direction of the local Lorentz force results in the correspondence relationship between the magnetic configurations associated with the flares and the magnetic chirality in the flare source regions. It is speculated that two-ribbon flares can be generally classified to these two magnetic configurations (Z-shaped or inverse Z-shaped) by magnetic chirality (left-handed or right-handed) in the flare source regions of ARs. This speculation needs to be validated by analyzing a large sample of solar flare events (both eruptive and non-eruptive) in the future work.

The analysis results of this study can be applied to the solar flare prediction studies and services \citep{2008AdSpR..42.1450H, 2012agos...30..117H, 2018HGSS....9...41H, 2008AdSpR..42.1464W, 2009RAA.....9..687W, 2018IAUS..335..243W}. The magnetic configuration of solar flares can also provide useful implications for analyzing the magnetic and flare activity properties of solar-type stars \citep{2015ApJS..221...18H, 2018IAUS..335....7H, 2018ApJS..236....7H, 2018IAUS..340..217H, 2016AcASn..57....9Y, 2017ChA&A..41...32Y, 2017ApJ...834..207M, 2018MNRAS.479L.139L, 2019NewA...66...31M, 2019ApJS..244...37G, 2020ApJS..247....9Z}.

\section*{Acknowledgments}

We thank the colleagues for valuable comments and discussions during the 15th Annual Meeting of the Asia Oceania Geosciences Society (AOGS2018) Session ST01 Flare Activity: Observation, Physics, and Forecasting and the 2018 ISEE/CICR International Workshop on Data-Driven Models of the Solar Progenitors of Space Weather and Space Climate. SDO is a mission of NASA's Living With a Star program. Hinode is a Japanese mission developed and launched by ISAS/JAXA, with NAOJ as domestic partner and NASA and STFC (UK) as international partners. It is operated by these agencies in co-operation with ESA and NSC (Norway). SOHO is a project of international cooperation between ESA and NASA. The vector magnetic field data of SDO-HMI were provided by the Joint Science Operations Center (JSOC; \url{http://jsoc.stanford.edu/}). The Level 2 data of Hinode-SP were provided by the Community Spectro-polarimetric Analysis Center (CSAC; \url{https://www2.hao.ucar.edu/csac}). SunPy \citep{2015CS&D....8a4009S, 2020ApJ...890...68S}, Astropy \citep{2013A&A...558A..33A, 2018AJ....156..123A}, NumPy \citep{2007CSE.....9c..10O}, Matplotlib \citep{2007CSE.....9...90H}, and IPython \citep{2007CSE.....9c..21P} were used in this research. This work belongs to a project supported by the Specialized Research Fund for Shandong Provincial Key Laboratory. The authors acknowledge the supports of the National Natural Science Foundation of China (NSFC; grants 11973059, 11761141002, 40890160, 40890161, and 10803011), the National Key R\&D Program of China (grants 2019YFA0405000 and 2014FY120300), the Astronomical Big Data Joint Research Center, co-founded by the National Astronomical Observatories, Chinese Academy of Sciences and the Alibaba Cloud, the Strategic Priority Research Program on Space Science, Chinese Academy of Sciences (grants XDA15052200 and XDB41000000), and the China Meteorological Administration (grant GYHY201106011).

%% The Appendices part is started with the command \appendix;
%% appendix sections are then done as normal sections
\appendix

\section{Deflection of the magnetic field at the magnetic connectivity breaking site before and after the two flares}  \label{sec:appendix}

\begin{figure}[htbp]
  \centering
  \includegraphics[scale=0.72]{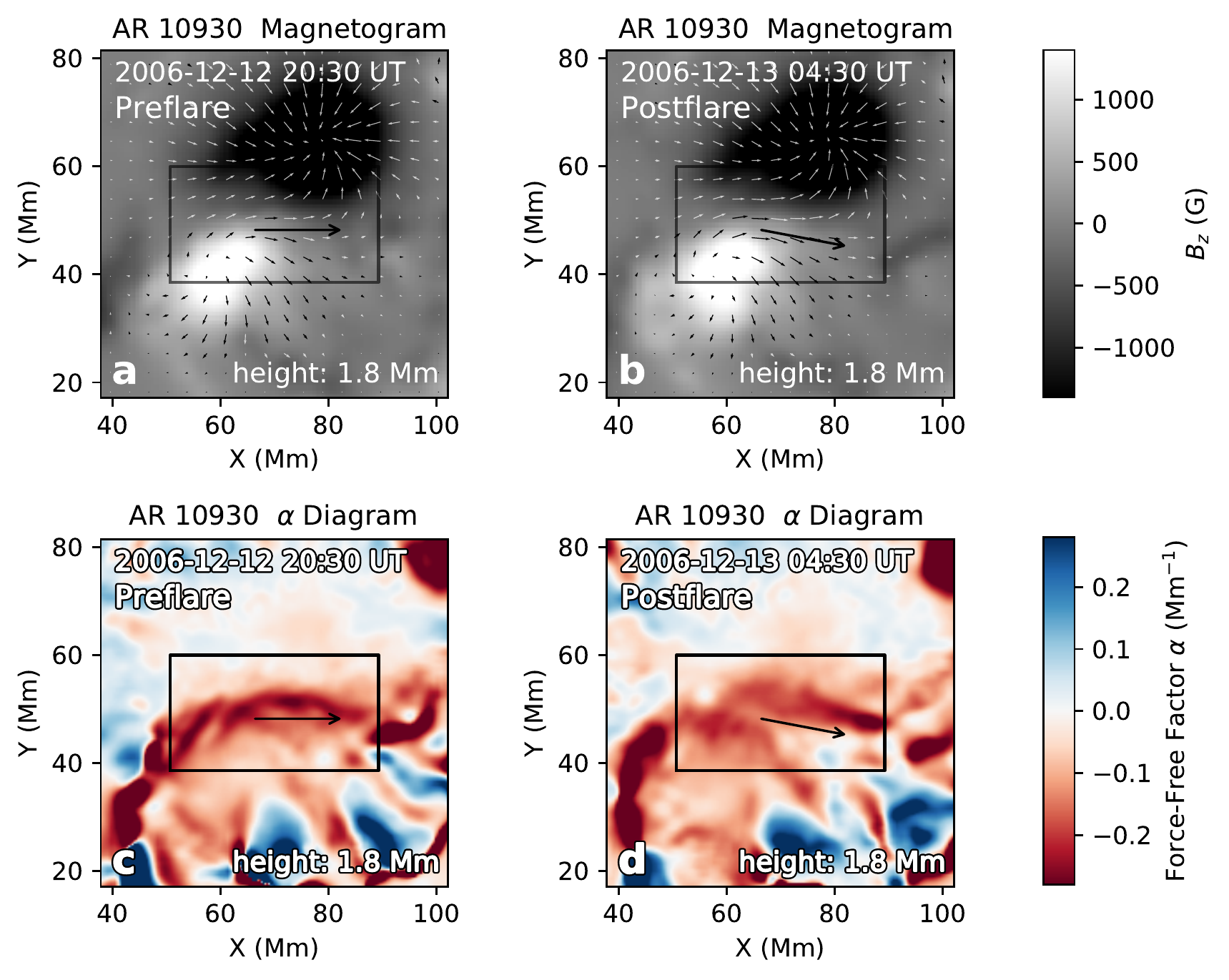}
  \caption{Diagram illustrating the deflection of the magnetic field at the magnetic connectivity breaking site before and after the flare in AR 10930. Top row: vector magnetograms at the height of 1.8 Mm in the central area of the AR before (left column) and after (right column) the flare; bottom row: $\alpha$ distribution maps at at the height of 1.8 Mm corresponding to the magnetograms. The black box in each panel indicates the main PIL zone of AR 10930 (same box position as in previous figures). The azimuths of the magnetic field at the magnetic connectivity breaking site before and after the flare are indicated by the long black arrows located near the center of each panel. The coordinates of the start point of the long arrows are $X=66.4$ Mm, $Y=48.2$ Mm, where the magnetic field azimuths are sampled. In the top panels, the gray scale images display the vertical component ($B_z$) of the magnetograms and the small arrows overlying the gray scale images represent the transverse component ($\vec{B}_t$) of the magnetograms. Note that the long black arrows belong to the arrow array of $\vec{B}_t$; they are enlarged to demonstrate the deflection of the magnetic field at the magnetic connectivity breaking site more clearly. \label{fig:fig8}}
\end{figure}

\begin{figure}[htbp]
  \centering
  \includegraphics[scale=0.72]{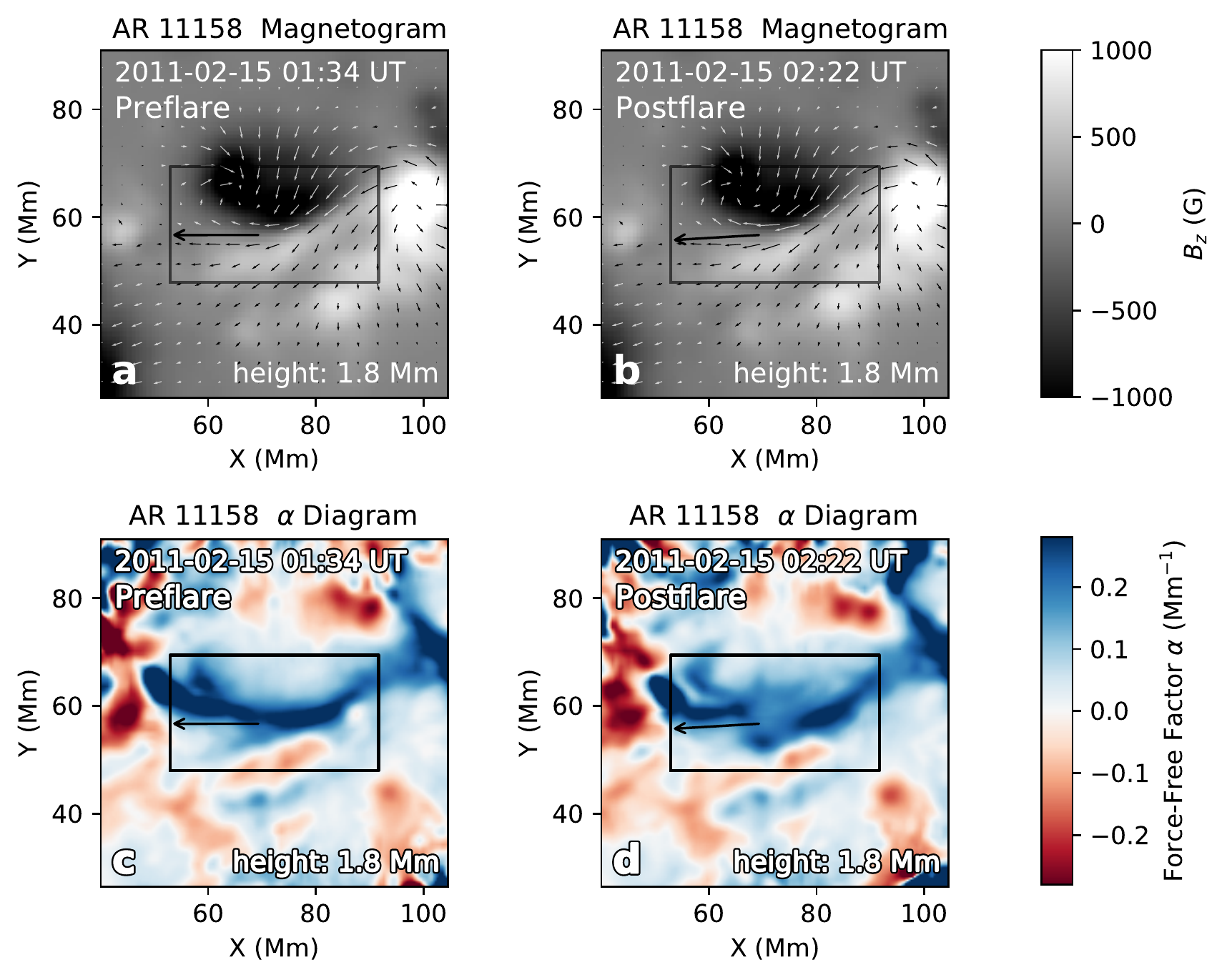}
  \caption{Same as Figure \ref{fig:fig8}, but for AR 11158. The magnetic field azimuths (indicated by the long black arrows) at the magnetic connectivity breaking site are sampled at $X=69.3$ Mm, $Y=56.7$ Mm (start point of the long arrows). \label{fig:fig9}}
\end{figure}

In Figures \ref{fig:fig8} and \ref{fig:fig9}, we display the magnetic field azimuths at the magnetic connectivity breaking site before and after the two flares in AR 10930 (see Figure \ref{fig:fig8}) and AR 11158 (see Figure \ref{fig:fig9}). In each figure, the top panels show the vector magnetograms at the height of 1.8 Mm in the central area of the AR before and after the flare; and the bottom panels show the corresponding $\alpha$ distribution maps at the same height. The azimuths of the magnetic field at the magnetic connectivity breaking site are indicated by the long black arrows located near the center of each panel.

Figures \ref{fig:fig8} and \ref{fig:fig9} illustrate the deflection of the magnetic field at the magnetic connectivity breaking site before and after the flares in both the ARs. For AR 10930, the magnetic field deflection is clockwise; and for AR 11158 the deflection is anticlockwise. From the bottom panels of Figures \ref{fig:fig8} and \ref{fig:fig9} ($\alpha$ distribution diagrams) it can be seen that, for both the flare events in the two ARs, the azimuth of the magnetic field at the magnetic connectivity breaking site is roughly aligned with the prominent magnetic connectivity before the flare and roughly aligned with the two branches of broken magnetic connectivity after the flare. This result is used in Section \ref{sec:discus} (Discussion) and Figure \ref{fig:fig6}.

It should be noticed that near the PIL, the strength of the transverse field in photosphere increases after flares \citep[e.g.,][]{2010ApJ...716L.195W, 2012ApJ...745L..17W, 2012ApJ...745L...4L}, which is interpreted by the back reaction effect of flare eruptions in literature \citep[e.g.,][]{2010ApJ...716L.195W, 2008ASPC..383..221H}. The breaking of the prominent magnetic connectivity and the deflection of the magnetic field at the magnetic connectivity breaking site discussed in this paper are different phenomena. They happen in low corona and above the photosphere, and are driven by the evolution of the photospheric magnetic field which acts as the bottom boundary condition of the corona system \citep{2014JGRA..119.3286H}.

\newpage
%% If you have bibdatabase file and want bibtex to generate the
%% bibitems, please use
%%
%\bibliographystyle{elsarticle-harv}
%\bibliography{reference}

%% else use the following coding to input the bibitems directly in the
%% TeX file.

%\begin{thebibliography}{00}

%% \bibitem[Author(year)]{label}
%% Text of bibliographic item

%\bibitem[ ()]{}

%\end{thebibliography}
\end{document}